\documentclass[a4paper,twoside,12pt]{article}
\input{obsjkt.sty}

\newcommand{\Msun}{\ensuremath{~{\rm M}_\odot}}                   
\newcommand{\Rsun}{\ensuremath{~{\rm R}_\odot}}                   
\newcommand{\rhosun}{\ensuremath{~\rho_\odot}}                    
\newcommand{\EBV}{\ensuremath{E(B\!-\!V)}}                        
\newcommand{\Grp}{\ensuremath{G_{\rm RP}}}                        
\newcommand{\degr}{\ensuremath{^\circ}}                           
\renewcommand{\kms}{~km~s$^{-1}$}                                 
\renewcommand{\cd}{~d$^{-1}$}                                     
\newcommand{\chir}{\ensuremath{\chi_\nu^{\,2}}}                   
\newcommand{\mc}[1]{\multicolumn{2}{c}{#1}}                       
\newcommand{\gaia}{\textit{Gaia}}                                 
\newcommand{\targ}{CV~Vel}
\newcommand{\targfull}{CV~Velorum}

\newcommand{\Msunnom}{\hbox{$\mathcal{M}^{\rm N}_\odot$}}
\newcommand{\Rsunnom}{\hbox{$\mathcal{R}^{\rm N}_\odot$}}
\newcommand{\Lsunnom}{\hbox{$\mathcal{L}^{\rm N}_\odot$}}

\usepackage{rotating}

\begin{document} 

\OBSheader{Rediscussion of eclipsing binaries: \targ}{J.\ Southworth}{2026 August}

\OBStitle{Rediscussion of eclipsing binaries. Paper 31. \\ The slowly-pulsating B-star system CV Velorum in the PLATO southern field}

\OBSauth{John Southworth}

\OBSinstone{Astrophysics Group, Keele University, Staffordshire, ST5 5BG, UK}


\OBSabstract{\targ\ is a detached eclipsing binary containing two B2.5\,V stars in a circular orbit of period 6.889~d. Both stars show line-profile variations arising from g-mode oscillations, characteristic of slowly-pulsating B-stars, and rotational axes misaligned with the orbital axis. We present the first photometric analysis of the system based on light curves from a space telescope. By combining our results with published spectroscopic observations we determine the stars' masses to be $6.065 \pm 0.015$\Msun\ and $5.950 \pm 0.012$\Msun, and their radii to be $4.094 \pm 0.056$\Rsun\ and $3.978 \pm 0.049$\Rsun. The precision of the radius measurements is limited by the pulsations in the light curves. We identify two confirmed (0.4886\cd\ and 0.3695\cd) and two candidate (0.7468\cd\ and 0.7048\cd) pulsation frequencies but cannot reliably identify further frequencies due to the short duration of the available light curves (approximately 50~d). \targ\ is in the first field to be observed by the PLATO satellite so in future may become the first EB containing an SPB star for which we have a light curve with a high photometric precision and a duration of over one year.}


\section*{Introduction}

Eclipsing binaries (EBs) are our primary source of directly measured properties of normal stars \cite{Andersen91aarv,Torres++10aarv} because their masses and radii can be measured to high precision and accuracy using only light and radial velocity (RV) curves \cite{Maxted+20mn}. Detached eclipsing binaries (dEBs) are of particular value as benchmarks against which the predictions of single-star evolutionary theory can be tested. The current work is part of a project \cite{Me20obs} to reanalyse known dEBs using published spectroscopy and new space-based light curves \cite{Me21univ}, for inclusion in the Detached Eclipsing Binary Catalogue \cite{Me15aspc} (DEBCat\footnote{\texttt{https://www.astro.keele.ac.uk/jkt/debcat/}}).

In this work we present an analysis of \targfull\ (Table~\ref{tab:info}), which consists of two B2.5\,V stars in an orbit of period 6.889~d. A rare -- possibly unique -- feature of \targ\ is that both stars belong to the class of slowly-pulsating B (SPB) stars. SPB stars are gravity-mode pulsators with masses of approximately 3\Msun\ to 8\Msun\ and lying within the main-sequence band \cite{WaelkensRufener85aa,Waelkens91aa}. They show multi-periodic g-mode pulsations of high radial order and frequencies normally from 0.25\cd\ to 1\cd, excited by the heat-engine mechanism \cite{Dziembowski++93mn}. SPB pulsations can be used to constrain interior properties such as the efficiencies of mixing and angular momentum transport \cite{Pedersen+21natas,Pedersen22apj}. 

Only a few SPB stars are known in dEBs: this list includes V539\,Ara \cite{Clausen96aa}, $\mu$\,Eri \cite{Jerzykiewicz+13mn}, AR\,Cas and AS\,Cam (papers in prep.), and a few others found in surveys \cite{MeBowman22mn,Cakirli+25mn}. The combination of pulsations and eclipses is expected to lead to new insights into the physics governing the evolution of massive stars \cite{MeBowman26araa} but light curves of sufficient duration (several years) to study SPB stars in EBs remain unavailable. This may soon change, because \targ\ is in the southern field of the forthcoming PLATO mission \cite{Rauer+25exa}. Here we present a photometric study of \targ\ using space-based photometry, measuring pulsation frequencies in this object for the first time.


\section*{\targfull}

\begin{table}[t]
\caption{\em Basic information on \targfull. 
The $BV$ magnitudes are each the mean of 151 individual measurements \cite{Hog+00aa} distributed approximately randomly in orbital phase. 
The $JHK_s$ magnitudes from 2MASS \cite{Cutri+03book} were obtained at an orbital phase of 0.76. \label{tab:info}}
\centering
\begin{tabular}{lll}
{\em Property}                            & {\em Value}                 & {\em Reference}                      \\[3pt]
Right ascension (J2000)                   & 09 00 37.99                 & \citenum{Gaia23aa}                   \\
Declination (J2000)                       & $-$51 33 20.1               & \citenum{Gaia23aa}                   \\
Henry Draper designation                  & HD 77464                    & \citenum{CannonPickering19anhar2}    \\
\textit{Tycho} designation                & TYC 8177-1750-1             & \citenum{Hog+00aa}                   \\
\textit{Gaia} DR3 designation             & 5323799125200766592         & \citenum{Gaia21aa}                   \\
\textit{Gaia} DR3 parallax (mas)          & $1.7652 \pm 0.0333$         & \citenum{Gaia21aa}                   \\          
TESS\ Input Catalog designation           & TIC 401238440               & \citenum{Stassun+19aj}               \\
$B$ magnitude                             & $6.507 \pm 0.012$           & \citenum{Hog+00aa}                   \\          
$V$ magnitude                             & $6.679 \pm 0.012$           & \citenum{Hog+00aa}                   \\          
$J$ magnitude                             & $6.963 \pm 0.024$           & \citenum{Cutri+03book}               \\
$H$ magnitude                             & $7.016 \pm 0.033$           & \citenum{Cutri+03book}               \\
$K_s$ magnitude                           & $7.061 \pm 0.020$           & \citenum{Cutri+03book}               \\
Spectral type                             & B2.5\,V + B2.5\,V           & \citenum{Andersen75aa}               \\[3pt]       
\end{tabular}
\end{table}



\targ\ was discovered to be eclipsing by van Houten in 1950 \cite{Vanhouten50anlei} in the course of a search for new variable stars using 18 photographic plates and a blink comparator. He found a period of 3.445~d and reported its spectral type to be B3.

Feast \cite{Feast54mn} discovered \targ\ to be a double-lined spectroscopic binary. He obtained the first spectroscopic orbit, based on 34 photographic spectra with reciprocal dispersions of 49 or 29 \AA~mm$^{-1}$, with a period of 6.892~d. This doubling of the orbital period is due to the primary and secondary eclipses being of very similar depth.

Gaposchkin \cite{Gaposchkin55mn} presented a photographic light curve. He confirmed Feast's period measurement and, using Feast's RVs, provided the first measurement of the masses and radii of the component stars.

Andersen \cite{Andersen75aa} presented a spectroscopic study of \targ\ based on 32 photographic spectra \cite{Andersen75aa2} covering 3700--4700~\AA\ with a reciprocal dispersion of 12.3~\AA~mm$^{-1}$. He found that ``the two components appear virtually indistinguishable in both spectral type and luminosity'' and adopted spectral types of B2.5\,V for both. The lines were ideal for RV measurement due to their strength and low rotational broadening (given as 28\kms\ $\pm$10\% for both components). Andersen measured RVs from 31 different absorption lines and demonstrated that line blending causes an underestimation of the RVs of the stars measured from H and diffuse He lines. His final spectroscopic orbit rested on RVs from a set of ten strong and unblended lines arising from C, Ca, Si, Mg and sharp He lines.

Clausen \& Gr{\o}nbech \cite{ClausenGronbech77aa} observed \targ\ using the 50\,cm Copenhagen telescope at ESO La Silla \cite{Gronbech++76aas}, obtaining complete coverage of the eclipses in the Str\"omgren $u$, $b$, $v$ and $y$ bands. They found the secondary eclipse to occur slightly before phase 0.5, indicating a small but non-zero eccentricity. They modelled the data using the {\sc wink} code \cite{Wood71aj} and determined the masses and radii of the component stars. A light ratio of $0.90 \pm 0.02$, obtained on request by Andersen from his spectra, was used to constrain the light-curve solution and obtain a determinate value of the ratio of the radii of the stars.

Yakut et al.\ \cite{Yakut++07aa} obtained 30 high-quality \'echelle spectra of \targ\ and used them to determine new spectroscopic orbits and physical properties of the stars\footnote{Note that Yakut et al.\ \cite{Yakut++07aa} switched the identities of the two stars \cite{Albrecht+14apj}. Their results quoted in the current work have been updated to account for this.}. They found the secondary star (hereafter star~B) to rotate faster than the primary (star~A): $V_{\rm A} \sin i = 31 \pm 2$\kms\ and $V_{\rm B} \sin i = 19 \pm 1$\kms. They also found both stars to show line-profile variations (stronger in star~B than star~A), with periods very uncertain but close to the orbital period. The physical properties of both stars are consistent with those of SPB stars. They found effective temperatures of $19000 \pm 500$~K and chemical abundances close to solar.

Finally, Albrecht et al.\ \cite{Albrecht+14apj} obtained spectra of \targ\ during both eclipses to study the Rossiter-McLaughlin effect in the system \cite{Rossiter24apj,McLaughlin24apj}. They found star~A to be misaligned with the orbit (sky-projected orbital obliquity $\beta = -52^\circ \pm 6^\circ$) and star~B to be possibly aligned ($\beta = 3^\circ \pm 7^\circ$). In this model the precession of the spin axis of star~A causes its $V\sin i$ to vary on a period of approximately 140~yr. The observed change in the rotational broadening of star~B suggests it too has a significant true orbital obliquity, and that the measurement of $\beta$ happened at a time when the sky-projection of the obliquity was low.


\section*{Photometric observations}


\targ\ has been observed by the Transiting Exoplanet Survey Satellite (TESS ref{Ricker+15jatis}) mission in eight sectors so far, with a range of cadences. Sectors 8 and 9 were observed at 1800\,s cadence; 35, 36, 89 and 90 at 600\,s cadence; and 62 and 63 at 200\,s cadence. We used the {\sc lightkurve} package \cite{Lightkurve18} to download the light curves for all sectors from the Mikulski Archive for Space Telescopes (MAST\footnote{\texttt{https://mast.stsci.edu/portal/Mashup/Clients/Mast/Portal.html}}). Light curves from the TESS Science Processing Center (SPOC \cite{Jenkins+16spie}) are available for sectors 62 and 63, whereas only Quick Look Pipeline (QLP \cite{Huang+20rnaas}) data are available for the other sectors. 

\begin{figure}[t] \centering \includegraphics[width=\textwidth]{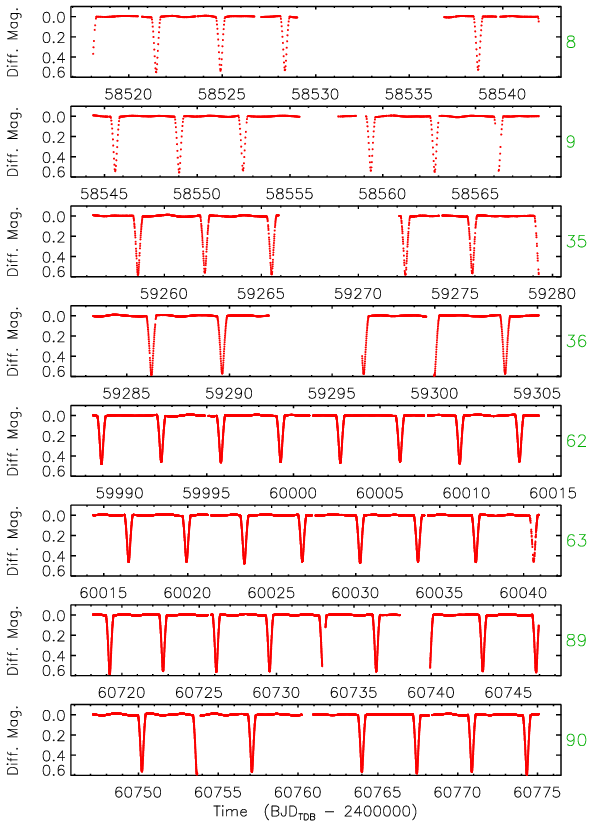} \\
\caption{\label{fig:time} Light curves of \targ\ from all eight TESS sectors. The flux 
measurements have been converted to magnitude units and the median subtracted.} \end{figure}

\clearpage

We converted the light curves from flux into differential magnitude, specified the ``hard'' option to reject data of lower quality, and subtracted median magnitude from each sector. The results are plotted in Fig.~\ref{fig:time}, where eclipses and pulsations are both visible. We based most of our analysis on sectors 62 and 63 because of the higher time resolution of the data.

\targ\ is close to the galactic plane so is in a relatively rich star field. A query of the \gaia\ DR3 database\footnote{\texttt{https://vizier.cds.unistra.fr/viz-bin/VizieR-3?-source=I/355/gaiadr3}} returns a total of 405 sources within 2~arcmin of \targ\ and thus potential contaminants of the TESS light curve. The brightest of these in the \gaia\ \Grp\ band, a red passband tolerably well matched to the TESS response function, is fainter by 4.46~mag. The fainter stars are relatively evenly spread so will be removed by the sky-subtraction step in the reduction of the TESS data. We therefore expect the amount of contaminating light in the TESS light curve of \targ\ to be small but not zero.


\section*{Light curve analysis}

We concentrated our analysis on the light curve formed by combining TESS sectors 62 and 63, totalling 21,746 data points, which have a higher time resolution than the rest of the data. The two stars are well-detached so are suitable for analysis with the {\sc jktebop}\footnote{\texttt{http://www.astro.keele.ac.uk/jkt/codes/jktebop.html}} code \cite{Me++04mn2,Me13aa}, for which we used version 45. We fitted for the orbital period ($P$), a reference time of primary minimum ($T_0$; when star~A is eclipsed by star~B) close to the midpoint of the data, the sum ($r_{\rm A}+r_{\rm B}$) and ratio ($k = r_{\rm B}/r_{\rm A}$) of the fractional radii, the central surface brightness ratio ($J$), third light ($L_3$), and orbital inclination ($i$). 

\begin{figure}[t] \centering \includegraphics[width=\textwidth]{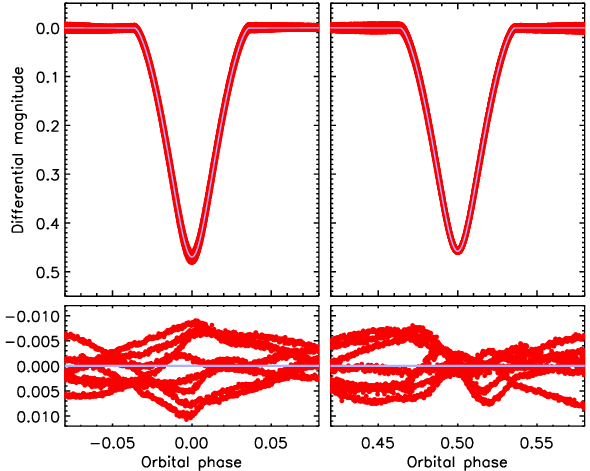} \\
\caption{\label{fig:phase} {\sc jktebop} best fit to the light curves of \targ\ from 
TESS sectors 62 and 63 for the primary eclipse (left panels) and secondary eclipse 
(right panels). The data are shown as filled red circles and the best fit as a light 
blue solid line. The residuals are shown on an enlarged scale in the lower panels, 
and are dominated by the pulsations.} \end{figure}

We included limb darkening (LD) using the power-2 law \cite{Hestroffer97aa,Maxted18aa,Me23obs2}, using the same coefficients for both stars due to their extreme similarity. The linear coefficient ($c$) was fitted and the nonlinear coefficient ($\alpha$) was fixed at a theoretical value \cite{ClaretSouthworth22aa,ClaretSouthworth23aa}. We tested for the presence of orbital eccentricity and found no substantive evidence, so assumed a circular orbit. The pulsations in the light curve are of much greater amplitude than caused by effects such as which LD law to use and whether to include eccentricity, so our ability to tease out these subtleties from the data is limited.

The best fit to the light curve is insensitive to $k$, as previously noted for this system \cite{ClausenGronbech77aa}. We followed Clausen \& Gr{\o}nbech in specifying a light ratio of $0.90 \pm 0.02$ measured by Andersen from two spectra of the system at quadrature. This light ratio was obtained at blue wavelengths but was applied directly in our {\sc jktebop} model of the much redder TESS data because the two stars have almost identical temperatures. We found this light ratio gives a significant improvement in the determinacy of our light-curve model. For illustration, in Table~\ref{tab:jktebop} we give the best-fitting parameters with and without the inclusion of this light ratio. The amount of third light is small and consistent with zero, as expected.

We were able to obtain a good fit to the eclipses (Fig.~\ref{fig:phase}). The residuals are dominated by the pulsations, which are at a maximum at the midpoint of primary eclipse and a minimum at the midpoint of secondary eclipse. As star~A dominates the light of the system during secondary eclipse and is mostly obscured during primary eclipse, we deduce that the dominant pulsations in the light curve arise from star~B. This agrees with previous observations of stronger line-profile variability in star~B than star~A \cite{Yakut++07aa,Albrecht+14apj}.

We determined the uncertainties in the fitted parameters using the residual-permutation algorithm implemented as Task 9 in {\sc jktebop} \cite{Me08mn}, thus treating the pulsations as correlated noise. There was no point conducting our usual Monte Carlo simulations as that algorithm assumes the data contain only the eclipse model and white noise. The best-fitting parameters and uncertainties are given in Table~\ref{tab:jktebop}. The uncertainties we find are driven by the pulsations, and are larger than those in recent publications \cite{Yakut++07aa,Albrecht+14apj}. Our orbital inclination measurement agrees well with that from Clausen \& Gr{\o}nbech \cite{ClausenGronbech77aa}, so there is no evidence for changes in this parameter as might be caused by precession of the rotational axes \cite{Albrecht+14apj}.

\begin{sidewaystable} \centering
\caption{\em \label{tab:jktebop} Photometric parameters measured using {\sc jktebop} 
from the TESS sector 62+63 light curve of \targ. The error bars are 1$\sigma$ and are 
calculated from residual-permutation simulations. The adopted solution is the first one.}
\begin{tabular}{lr@{\,$\pm$\,}lr@{\,$\pm$\,}lr@{\,$\pm$\,}l}
Parameter & \mc{~~$\ell_{\rm B}/\ell_{\rm A}$ constrained} & \mc{$\ell_{\rm B}/\ell_{\rm A}$ not constrained} & \mc{With 11 frequencies removed} \\[3pt]
{\it Fitted parameters:} \\
Orbital period (d)                        &       6.889466 & 0.000056 &       6.889466 & 0.000056 &         6.889488 & 0.000027 \\
Time of primary eclipse (BJD$_{\rm TDB}$) & 2460009.59623  & 0.00011  & 2460009.59623  & 0.00011  & ~~2460009.59618  & 0.00011  \\
Orbital inclination (\degr)               &      86.652    & 0.012    &      86.636    & 0.051    &        86.645    & 0.017    \\
Sum of the fractional radii               &       0.2313   & 0.0009   &       0.2315   & 0.0008   &         0.2319   & 0.0008   \\
Ratio of the radii                        &       0.971    & 0.024    &       1.016    & 0.091    &         1.009    & 0.018    \\
Central surface brightness ratio          &       0.9793   & 0.0048   &       0.9792   & 0.0048   &         0.9791   & 0.0014   \\
Third light                               &    $-$0.0023   & 0.0073   &       0.0000   & 0.0065   &         0.0036   & 0.0065   \\
LD coefficient $c$                        &       0.46     & 0.11     &       0.49     & 0.07     &         0.53     & 0.08     \\
LD coefficient $\alpha$                   &  \mc{~~0.3591 (fixed)}    &  \mc{~~0.3591 (fixed)}    &    \mc{~~0.3591 (fixed)}    \\
{\it Derived parameters:} \\                                                                                
Fractional radius of star~A               &       0.1173   & 0.0016   &       0.1148   & 0.0072   &         0.1154   & 0.0012   \\
Fractional radius of star~B               &       0.1140   & 0.0014   &       0.1167   & 0.0051   &         0.1164   & 0.0012   \\
Light ratio $\ell_{\rm B}/\ell_{\rm A}$   &       0.924    & 0.042    &       1.01     & 0.17     &         0.996    & 0.035    \\[3pt]
\end{tabular}
\end{sidewaystable}  

We considered adding the TESS data from sectors 35, 36, 89 and 90 to this analysis, to average out the effect of the pulsations and potentially allow more precise measurements of the properties in Table~\ref{tab:jktebop}. However, these data show much deeper eclipses (0.59~mag) than those from sectors 62 and 63 (0.48~mag) so it is not straightforward to combine them. Previous light curves of \targ\ show eclipse depths in agreement with those from sectors 62 and 63, so we conclude that the different data reduction procedures for QLP versus SPOC give an incorrect light curve scaling. We have seen this before for other dEBs.

We then attempted to lower the error bars in Table~\ref{tab:jktebop} by removing the majority of the pulsational effects from the light curve. We used {\sc period04} (see below) to fit and subtract 11 frequencies which plausibly exist, then refitted the light curve with {\sc jktebop}. This gave a much improved fit but only a small change in some of the fitted light curve parameters. The solution prefers a higher light ratio than that measured by Andersen, giving a ratio of the radii that is closer to unity and less consistent with expectations from stellar evolutionary theory. This new fit is also given in Table~\ref{tab:jktebop} for reference, but we did not use its results in the remainder of our analysis.


\section*{Orbital ephemeris}

\begin{table} \centering
\caption{\em Times of mid-eclipse for \targ\ and their residuals versus the fitted ephemeris. \label{tab:tmin}}
\setlength{\tabcolsep}{10pt}
\begin{tabular}{rrrrr}
{\em Orbital} & {\em Eclipse time}  & {\em Uncertainty} & {\em Residual} & {\em TESS}   \\
{\em cycle}   & {\em (BJD$_{TDB}$)} & {\em (d)}         & {\em (d)}      & {\em sectors} \\[3pt]

$-$212.0 & 2458549.02243 & 0.00013 & $-0.00018$ & 08, 09 \\                                       
$-$105.0 & 2459286.19909 & 0.00009 & $ 0.00015$ & 35, 36 \\                                       
     0.0 & 2460009.59623 & 0.00011 & $-0.00006$ & 62, 63 \\                                       
   107.0 & 2460746.77262 & 0.00006 & $-0.00001$ & 89, 90 \\                                       
\end{tabular}
\end{table}

We calculated an improved orbital ephemeris based only on the TESS data. We fitted each pair of adjacent TESS sectors with {\sc jktebop} to obtain one effective time of primary eclipse per pair of sectors. These are given in Table~\ref{tab:tmin}. A straight-line fit to these data returns the following linear ephemeris:
\begin{equation}
\mbox{Min~I} = {\rm BJD}_{\rm TDB}~ 2460009.59629 (6) + 6.8894985 (4) E
\end{equation}
where $E$ is the number of cycles since the reference time of minimum and the bracketed quantities indicate the uncertainty in the final digit of the previous number. The scatter around the best fit is slightly larger than the error bars suggest, with a reduced $\chi^2$ of $\chir = 1.2$, so the uncertainties in the ephemeris have been multiplied by $\sqrt{\chir}$ to account for this. There is no evidence for a change in the orbital period based on these data, albeit over a small time interval. Further historical times of minimum exist, as do more recent survey data, which could be used for a more extensive search for orbital period changes.

\section*{Physical properties and distance to \targ}

We determined the physical properties of \targ\ using our results from the light curve analysis above ($r_{\rm A}$, $r_{\rm B}$, $i$ and $P$ from Table~\ref{tab:jktebop}) and the {\sc jktabsdim} code \cite{Me++05aa}. To these we added the velocity amplitudes of $K_{\rm A} = 126.69 \pm 0.11$\kms\ and $K_{\rm B} = 129.15 \pm 0.11$\kms\ from Albrecht et al.\ \cite{Albrecht+14apj}, adding the quoted statistical and systematic errors in quadrature to give a single error bar for each quantity. The temperatures of the stars were taken from Torres et al.\ \cite{Torres++10aarv}, which are representative of most values quoted in the literature. The results are given in Table~\ref{tab:absdim}.

\begin{table} \centering
\caption{\em Physical properties of \targ\ defined using the nominal solar units 
given by IAU 2015 Resolution B3 (ref.~\citenum{Prsa+16aj}). \label{tab:absdim}}
\begin{tabular}{lr@{~$\pm$~}lr@{~$\pm$~}l}
{\em Parameter}        & \multicolumn{2}{c}{\em Star A} & \multicolumn{2}{c}{\em Star B}     \\[3pt]
Mass ratio   $M_{\rm B}/M_{\rm A}$          &  \multicolumn{4}{c}{$0.9810 \pm 0.0014$}        \\
Semimajor axis of relative orbit (\Rsunnom) &  \multicolumn{4}{c}{$34.899 \pm 0.025$}         \\
Mass (\Msunnom)                             &   6.065  & 0.015       &   5.950  & 0.012       \\
Radius (\Rsunnom)                           &   4.094  & 0.056       &   3.978  & 0.049       \\
Surface gravity ($\log$[cgs])               &   3.997  & 0.012       &   4.013  & 0.012       \\
Density ($\!\!$\rhosun)                     &   0.0884 & 0.0036      &   0.0945 & 0.0035      \\
Synchronous rotational velocity ($\!\!$\kms)&  30.06   & 0.41        &  29.22   & 0.36        \\
Effective temperature (K)                   &  18100   & 500         &  17900   & 500         \\
Luminosity $\log(L/\Lsunnom)$               &  3.210   & 0.049       &   3.166  & 0.049       \\
$M_{\rm bol}$ (mag)                         &$-$3.28   & 0.12        &$-$3.17   & 0.13        \\
Interstellar reddening \EBV\ (mag)          &  \multicolumn{4}{c}{$0.06 \pm 0.02$}	          \\
Distance (pc)                               &  \multicolumn{4}{c}{$568.4 \pm 9.5$}            \\[3pt]       
\end{tabular}
\end{table}


To measure the distance to the system we used the parameters above, the $BV$ magnitudes from Tycho \cite{Hog+00aa} and $JHK_s$ magnitudes from 2MASS \cite{Cutri+03book} given in Table~\ref{tab:info}, and bolometric corrections from Girardi et al.\ \cite{Girardi+02aa}. A reddening of $\EBV = 0.06 \pm 0.02$~mag brought the distances in the $BV$ passband into alignment with those in $JHK_s$. This led to a final distance measurement in the $K_s$ band (which is least affected by reddening) of $568 \pm 10$~pc, in excellent agreement with the distance of $567 \pm 11$~pc from inversion of the \gaia\ DR3 parallax.

The age of the system was estimated by comparing the measured masses, radii and temperatures to theoretical predictions from the {\sc parsec} 1.2 evolutionary models \cite{Bressan+12mn}. We found an essentially perfect fit for an age of $40 \pm 1$~Myr assuming a solar metal abundance fraction by mass of $Z=0.017$. A lower metal abundance of $Z=0.014$ gives a slightly worse fit (but still well within the error bars) for an age older by 3~Myr. Similarly, a higher metal abundance of $Z=0.020$ leads to a preference for an age younger by only 1~Myr.

\section*{Pulsation analysis}

\begin{figure}[t] \centering \includegraphics[width=\textwidth]{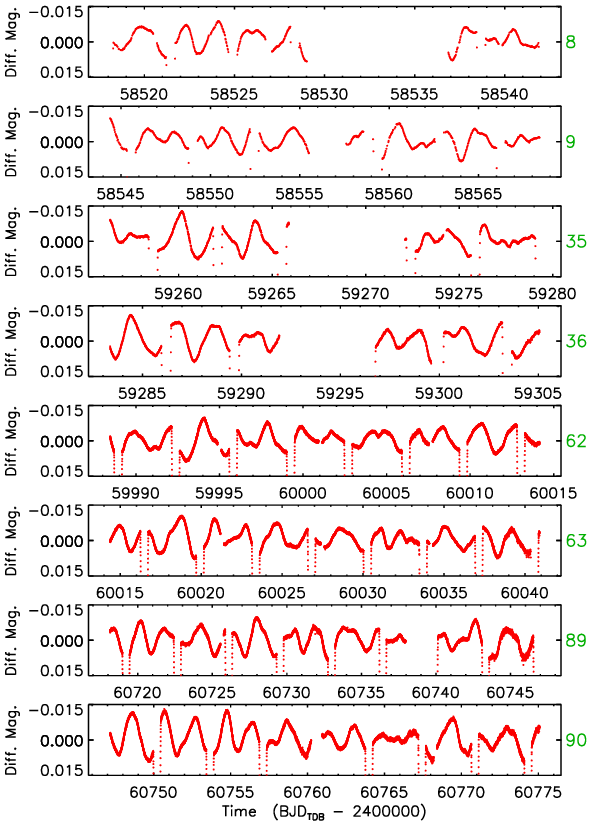} \\
\caption{\label{fig:close} Same as Fig.~\ref{fig:time} but with the 
ordinate axis enlarged to make the pulsations clearer.} \end{figure}

\begin{sidewaysfigure} \centering
\includegraphics[width=\textwidth]{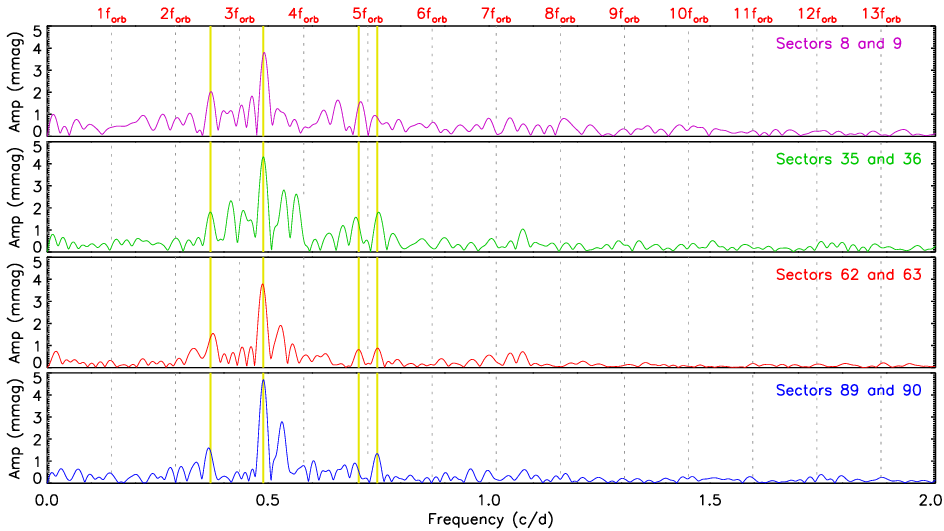}
\caption{\label{fig:freq} Frequency spectra of \targ\ from the four pairs of 
TESS sectors (labelled in the top-right of each panel). The frequency spectra
are plotted with coloured lines. Multiples of the orbital frequency are plotted
with grey dashed lines and annotated at the top of the plot. The four frequencies 
discussed in the text are indicated with thick yellow lines.} \end{sidewaysfigure}

The TESS light curve of \targ\ shows clear evidence for multi-periodic pulsations (see Fig.~\ref{fig:close}), as expected from the prior detections of line-profile variations in both stars \cite{Yakut++07aa,Albrecht+14apj}. We combined the eight TESS sectors into four pairs, fitted and subtracted a {\sc jktebop} model, then calculated frequency spectra for each pair of sectors using version 1.2.0 of the {\sc period04} code \cite{LenzBreger05coast}. These are shown in Fig.~\ref{fig:freq}.

\clearpage


All four pairs of sectors shows the strong detection of an oscillation with a frequency of $f_1 = 0.4886$\cd, an amplitude of 3.8~mmag, and a signal-to-noise ratio\footnote{The S/N is the ratio of the height of the peak to the mean signal within an interval of 2\cd\ centred on the peak.} (S/N) of 9.5. In the current paragraph we quote results from the data from sectors 62 and 63 as it has the lowest noise of the four pairs of sectors. A second frequency at $f_2 = 0.3695$\cd\ and amplitude 1.6~mmag is detected with a S/N of only 3.6, lower than needed for confirmed signals \cite{BaranKoen21aca}, but its presence in all four sector pairs supports its reality. Two further candidate frequencies, $f_3 = 0.7468$\cd\ (amplitude 0.9~mmag) and $f_4 = 0.7048$\cd\ (amplitude 0.7~mmag) are detected with S/N values of 2.9 and 2.7, respectively. $f_3$ and $f_4$ are each visible in three of the four sector pairs so are likely real but not confirmable with the currently-available data. We did not see any peaks in the frequency spectrum near multiples of the orbital frequency. We also checked and found no significant frequencies from 1.1\cd\ out to the Nyquist limit of 215\cd.

The confirmed and tentative frequency detections all lie at low frequencies, in a range characteristic of the SPB pulsators \cite{Waelkens91aa,MeBowman22mn}. The properties of both stars are also inside and close to the hotter boundary for which SPB pulsations are expected \cite{Dziembowski++93mn}. As both stars show line-profile variations, we conclude that they are both SPB pulsators. We defer a detailed frequency analysis until light curves of much longer duration are available in future.



\section*{Summary and conclusions}

\targ\ is a dEB containing two almost-identical B2.5\,V stars in a circular orbit of period 6.889~d. We have determined the masses and radii of the stars based on published spectroscopic studies and an analysis of light curves from the TESS mission. The pulsations in the light curve prevented us from determining the radii of the stars to high precision. We found a distance in good agreement with the parallax from \gaia\ DR3, and an age of 40~Myr from comparison with theoretical models. The rotational axes of both stars are misaligned with the orbital axis, and are precessing with periods of centuries. Due to this, the projected rotational velocities of the stars change over time.

Both stars are known to show line-profile variations indicative of pulsations of the SPB type. We have presented the first photometric detection of SPB pulsations and measured two certain and two tentative frequencies in the range 0.37\cd\ to 0.75\cd. Additional variability is visible in the frequency spectrum but no further frequencies are reliably detected in the available data. \targ\ is in the southern field (LOPS2 \cite{Nascimbeni+22aa,Nascimbeni+25aa}) of the PLATO mission \cite{Rauer+25exa}, which is scheduled for launch in early 2027 and will observe this field for (at least) the first two years of its mission. Observations of \targ\ from PLATO will allow a much more detailed analysis of the radii and pulsation frequencies of the star, and we will be proposing it as a high-priority target in the PLATO Guest Observer programme\footnote{\texttt{https://www.cosmos.esa.int/web/plato/guest-observers}}.


\section*{Acknowledgements}

We thank the anonymous referee for a helpful report.
We acknowledge support from STFC under grant number ST/Y002563/1.
This paper includes data collected by the TESS\ mission and obtained from the MAST data archive at the Space Telescope Science Institute (STScI). Funding for the TESS mission is provided by the NASA's Science Mission Directorate. STScI is operated by the Association of Universities for Research in Astronomy, Inc., under NASA contract NAS 5–26555.
This paper includes observations made with the Isaac Newton Telescope operated on the island of La Palma by the Isaac Newton Group of Telescopes in the Spanish Observatorio del Roque de los Muchachos of the Instituto de Astrof\'{\i}sica de Canarias.
This work has made use of data from the European Space Agency (ESA) mission {\it Gaia}\footnote{\texttt{https://www.cosmos.esa.int/gaia}}, processed by the {\it Gaia} Data Processing and Analysis Consortium (DPAC\footnote{\texttt{https://www.cosmos.esa.int/web/gaia/dpac/consortium}}). Funding for the DPAC has been provided by national institutions, in particular the institutions participating in the {\it Gaia} Multilateral Agreement.
The following resources were used in the course of this work: the NASA Astrophysics Data System; the SIMBAD database operated at CDS, Strasbourg, France; and the ar$\chi$iv scientific paper preprint service operated by Cornell University.



\begin{thebibliography}{10}
\newcommand{\enquote}[1]{`(#1)'}

\bibitem{Andersen91aarv}
J.~{Andersen}, \textit{A\&ARv}, \textbf{3}, 91, 1991.

\bibitem{Torres++10aarv}
G.~{Torres}, J.~{Andersen} \& A.~{Gim{\'e}nez}, \textit{A\&ARv}, \textbf{18},
  67, 2010.

\bibitem{Maxted+20mn}
P.~F.~L. {Maxted} \textit{et~al.}, \textit{MNRAS}, \textbf{498}, 332, 2020.

\bibitem{Me20obs}
J.~{Southworth}, \textit{The Observatory}, \textbf{140}, 247, 2020.

\bibitem{Me21univ}
J.~{Southworth}, \textit{Universe}, \textbf{7}, 369, 2021.

\bibitem{Me15aspc}
J.~{Southworth}, in \textit{Living Together: Planets, Host Stars and Binaries}
  (S.~M. {Rucinski}, G.~{Torres} \& M.~{Zejda}, eds.), 2015,
  \textit{Astronomical Society of the Pacific Conference Series}, vol. 496, p.
  321.

\bibitem{WaelkensRufener85aa}
C.~{Waelkens} \& F.~{Rufener}, \textit{A\&A}, \textbf{152}, 6, 1985.

\bibitem{Waelkens91aa}
C.~{Waelkens}, \textit{A\&A}, \textbf{246}, 453, 1991.

\bibitem{Dziembowski++93mn}
W.~A. {Dziembowski}, P.~{Moskalik} \& A.~A. {Pamyatnykh}, \textit{MNRAS},
  \textbf{265}, 588, 1993.

\bibitem{Pedersen+21natas}
M.~G. {Pedersen} \textit{et~al.}, \textit{Nature Astronomy}, \textbf{5}, 715,
  2021.

\bibitem{Pedersen22apj}
M.~G. {Pedersen}, \textit{ApJ}, \textbf{940}, 49, 2022.

\bibitem{Clausen96aa}
J.~V. {Clausen}, \textit{A\&A}, \textbf{308}, 151, 1996.

\bibitem{Jerzykiewicz+13mn}
M.~{Jerzykiewicz} \textit{et~al.}, \textit{MNRAS}, \textbf{432}, 1032, 2013.

\bibitem{MeBowman22mn}
J.~{Southworth} \& D.~M. {Bowman}, \textit{MNRAS}, \textbf{513}, 3191, 2022.

\bibitem{Cakirli+25mn}
{\"O}.~{{\c{C}}ak{\i}rl{\i}}, B.~{Hoyman} \& O.~{{\"O}zdarcan}, \textit{MNRAS},
  \textbf{544}, 607, 2025.

\bibitem{MeBowman26araa}
J.~{Southworth} \& D.~{Bowman}, \textit{ARA\&A, in press,
  \texttt{arXiv:2509.08426}}, 2026.

\bibitem{Rauer+25exa}
H.~{Rauer} \textit{et~al.}, \textit{Experimental Astronomy}, \textbf{59}, 26,
  2025.

\bibitem{Hog+00aa}
E.~{H{\o}g} \textit{et~al.}, \textit{A\&A}, \textbf{355}, L27, 2000.

\bibitem{Cutri+03book}
R.~M. {Cutri} \textit{et~al.}, \textit{{2MASS All Sky Catalogue of Point
  Sources}} (The IRSA 2MASS All-Sky Point Source Catalogue, NASA/IPAC Infrared
  Science Archive, Caltech, US), 2003.

\bibitem{Gaia23aa}
{Gaia Collaboration}, \textit{A\&A}, \textbf{674}, A1, 2023.

\bibitem{CannonPickering19anhar2}
A.~J. {Cannon} \& E.~C. {Pickering}, \textit{Annals of Harvard College
  Observatory}, \textbf{94}, 1, 1919.

\bibitem{Gaia21aa}
{Gaia Collaboration}, \textit{A\&A}, \textbf{649}, A1, 2021.

\bibitem{Stassun+19aj}
K.~G. {Stassun} \textit{et~al.}, \textit{AJ}, \textbf{158}, 138, 2019.

\bibitem{Andersen75aa}
J.~{Andersen}, \textit{A\&A}, \textbf{44}, 355, 1975.

\bibitem{Vanhouten50anlei}
C.~J. {van Houten}, \textit{Annalen van de Sterrewacht te Leiden}, \textbf{20},
  223, 1950.

\bibitem{Feast54mn}
M.~W. {Feast}, \textit{MNRAS}, \textbf{114}, 246, 1954.

\bibitem{Gaposchkin55mn}
S.~{Gaposchkin}, \textit{MNRAS}, \textbf{115}, 391, 1955.

\bibitem{Andersen75aa2}
J.~{Andersen}, \textit{A\&A}, \textbf{44}, 445, 1975.

\bibitem{ClausenGronbech77aa}
J.~V. {Clausen} \& B.~{Gr{\o}nbech}, \textit{A\&A}, \textbf{58}, 131, 1977.

\bibitem{Gronbech++76aas}
B.~{Gr{\o}nbech}, E.~H. {Olsen} \& B.~{Str{\"o}mgren}, \textit{A\&AS},
  \textbf{26}, 155, 1976.

\bibitem{Wood71aj}
D.~B. {Wood}, \textit{AJ}, \textbf{76}, 701, 1971.

\bibitem{Yakut++07aa}
K.~{Yakut}, C.~{Aerts} \& T.~{Morel}, \textit{A\&A}, \textbf{467}, 647, 2007.

\bibitem{Albrecht+14apj}
S.~{Albrecht} \textit{et~al.}, \textit{ApJ}, \textbf{785}, 83, 2014.

\bibitem{Rossiter24apj}
R.~A. {Rossiter}, \textit{ApJ}, \textbf{60}, 15, 1924.

\bibitem{McLaughlin24apj}
D.~B. {McLaughlin}, \textit{ApJ}, \textbf{60}, 22, 1924.

\bibitem{Lightkurve18}
{Lightkurve Collaboration}, \enquote{{Lightkurve: Kepler and TESS time series
  analysis in Python}}, Astrophysics Source Code Library, 2018.

\bibitem{Jenkins+16spie}
J.~M. {Jenkins} \textit{et~al.}, in \textit{Proc.\ SPIE}, 2016, \textit{Society
  of Photo-Optical Instrumentation Engineers (SPIE) Conference Series}, vol.
  9913, p. 99133E.

\bibitem{Huang+20rnaas}
C.~X. {Huang} \textit{et~al.}, \textit{RNAAS}, \textbf{4}, 204, 2020.

\bibitem{Me++04mn2}
J.~{Southworth}, P.~F.~L. {Maxted} \& B.~{Smalley}, \textit{MNRAS},
  \textbf{351}, 1277, 2004.

\bibitem{Me13aa}
J.~{Southworth}, \textit{A\&A}, \textbf{557}, A119, 2013.

\bibitem{Hestroffer97aa}
D.~{Hestroffer}, \textit{A\&A}, \textbf{327}, 199, 1997.

\bibitem{Maxted18aa}
P.~F.~L. {Maxted}, \textit{A\&A}, \textbf{616}, A39, 2018.

\bibitem{Me23obs2}
J.~{Southworth}, \textit{The Observatory}, \textbf{143}, 71, 2023.

\bibitem{ClaretSouthworth22aa}
A.~{Claret} \& J.~{Southworth}, \textit{A\&A}, \textbf{664}, A128, 2022.

\bibitem{ClaretSouthworth23aa}
A.~{Claret} \& J.~{Southworth}, \textit{A\&A}, \textbf{674}, A63, 2023.

\bibitem{Me08mn}
J.~{Southworth}, \textit{MNRAS}, \textbf{386}, 1644, 2008.

\bibitem{Me++05aa}
J.~{Southworth}, P.~F.~L. {Maxted} \& B.~{Smalley}, \textit{A\&A},
  \textbf{429}, 645, 2005.

\bibitem{Prsa+16aj}
A.~{Pr{\v s}a} \textit{et~al.}, \textit{AJ}, \textbf{152}, 41, 2016.

\bibitem{Girardi+02aa}
L.~{Girardi} \textit{et~al.}, \textit{A\&A}, \textbf{391}, 195, 2002.

\bibitem{Bressan+12mn}
A.~{Bressan} \textit{et~al.}, \textit{MNRAS}, \textbf{427}, 127, 2012.

\bibitem{LenzBreger05coast}
P.~{Lenz} \& M.~{Breger}, \textit{Communications in Asteroseismology},
  \textbf{146}, 53, 2005.

\bibitem{BaranKoen21aca}
A.~S. {Baran} \& C.~{Koen}, \textit{AcA}, \textbf{71}, 113, 2021.

\bibitem{Nascimbeni+22aa}
V.~{Nascimbeni} \textit{et~al.}, \textit{A\&A}, \textbf{658}, A31, 2022.

\bibitem{Nascimbeni+25aa}
V.~{Nascimbeni} \textit{et~al.}, \textit{A\&A}, \textbf{694}, A313, 2025.

\end{thebibliography}

\end{document}